\documentclass[aps,prl,twocolumn,floatfix]{revtex4}

\usepackage{amsmath,amssymb,bm}
\usepackage{epsfig}
\usepackage{graphics}

\begin{document}

\bibliographystyle{prsty}

\title
{Berry Phase and Pseudospin Winding Number in Bilayer Graphene}
\author{Cheol-Hwan Park$^{1,2}$}
\email{chpark77@mit.edu}
\author{Nicola Marzari$^{2,3}$}
\affiliation{
$^1$Department of Materials Science and Engineering, Massachusetts
Institute of Technology, Cambridge, Massachusetts 02139, USA\\
$^2$Department of Materials, University of Oxford, Oxford OX1 3PH, UK\\
$^3$Theory and Simulation of Materials, \'Ecole Polytechnique F\'ed\'erale de Lausanne, 1015 Lausanne, Switzerland}
\date{\today}

\begin{abstract}
Ever since the novel quantum Hall effect in bilayer graphene was
discovered, and explained by a Berry phase of $2\pi$
[K. S. Novoselov {\it et al}., ``Unconventional quantum Hall
effect and Berry's phase of $2\pi$ in bilayer graphene'', Nature Phys. {\bf 2}, 177 (2006)],
it has been widely accepted that the low-energy electronic
wavefunction in this system is described by a non-trivial Berry
phase of $2\pi$, different from the zero phase of a conventional two-dimensional
electron gas.
{Here, we show that (i) the relevant Berry phase for bilayer graphene is
{not different from}
that for a conventional two-dimensional electron gas
{(as expected, given that Berry phase is only meaningful modulo $2\pi$)}}
and that (ii) what is actually observed in the quantum Hall measurements is not
the absolute value of the Berry phase but the pseudospin winding number.
\end{abstract}
\maketitle

\section{I. Introduction}

\begin{figure}
  \includegraphics[width=1.0\columnwidth]{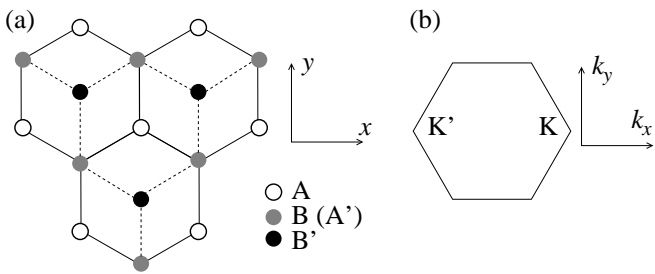}
\caption{
(a) Schematic representation of graphene and bilayer graphene.
For graphene, the carbon atoms belong only to the two A and B sublattices.
(b) The Brillouin zone of graphene and bilayer graphene.
    The positions of the K and K$'$ points are
    $\left(\frac{4\pi}{3a},0\right)$ and
    $\left(-\frac{4\pi}{3a},0\right)$, respectively, where $a$ is the
    lattice parameter.
}
\label{Fig1}
\end{figure}

A conventional two-dimensional electron gas (2DEG) under a high
magnetic field and low temperature shows a quantized Hall conductance
$\sigma_{xy}=n\, e^2/h$ per spin degree of freedom where $n$ is an
integer, $e$ the charge of an electron, and $h$ Planck's
constant~\cite{PhysRevLett.45.494}.
For graphene, it has been predicted that the
Hall conductance per spin and valley degrees of freedom is
$\sigma_{xy}=(n+1/2)\,e^2/h$~\cite{PhysRevB.65.245420,PhysRevLett.95.146801,peres_electronic_2006},
and this prediction has then been proven by experiments performed
on mechanically exfoliated
samples~\cite{novoselov:2005Nat_Graphene_QHE,zhang:2005Nat_Graphene_QHE}.
Such half-integer quantum Hall effect is the most decisive evidence
that the sample is actually a single atomic layer of carbon atoms
effectively decoupled from the substrate,
and is a direct manifestation of a non-trivial Berry phase of
$\pi$ in the electron
wavefunction~\cite{novoselov:2005Nat_Graphene_QHE,zhang:2005Nat_Graphene_QHE}.
(This half-integer quantum Hall effect is different from
the fractional quantum Hall effect
in graphene~\cite{du_fractional_2009,bolotin_observation_2009,ghahari_measurement_2011,dean_multicomponent_2010}.)
Half-integer quantum Hall effect has been observed in graphene epitaxially
grown on the carbon-rich surface~\cite{wu_half_2009}
and the silicon-rich
surface~\cite{shen_observation_2009,tzalenchuk_towards_2010,jobst_quantum_2010}
of silicon carbide
or in graphene grown by chemical vapor deposition~\cite{kim_large-scale_2009}.

  \begin{figure*}
  \includegraphics[width=2.0\columnwidth]{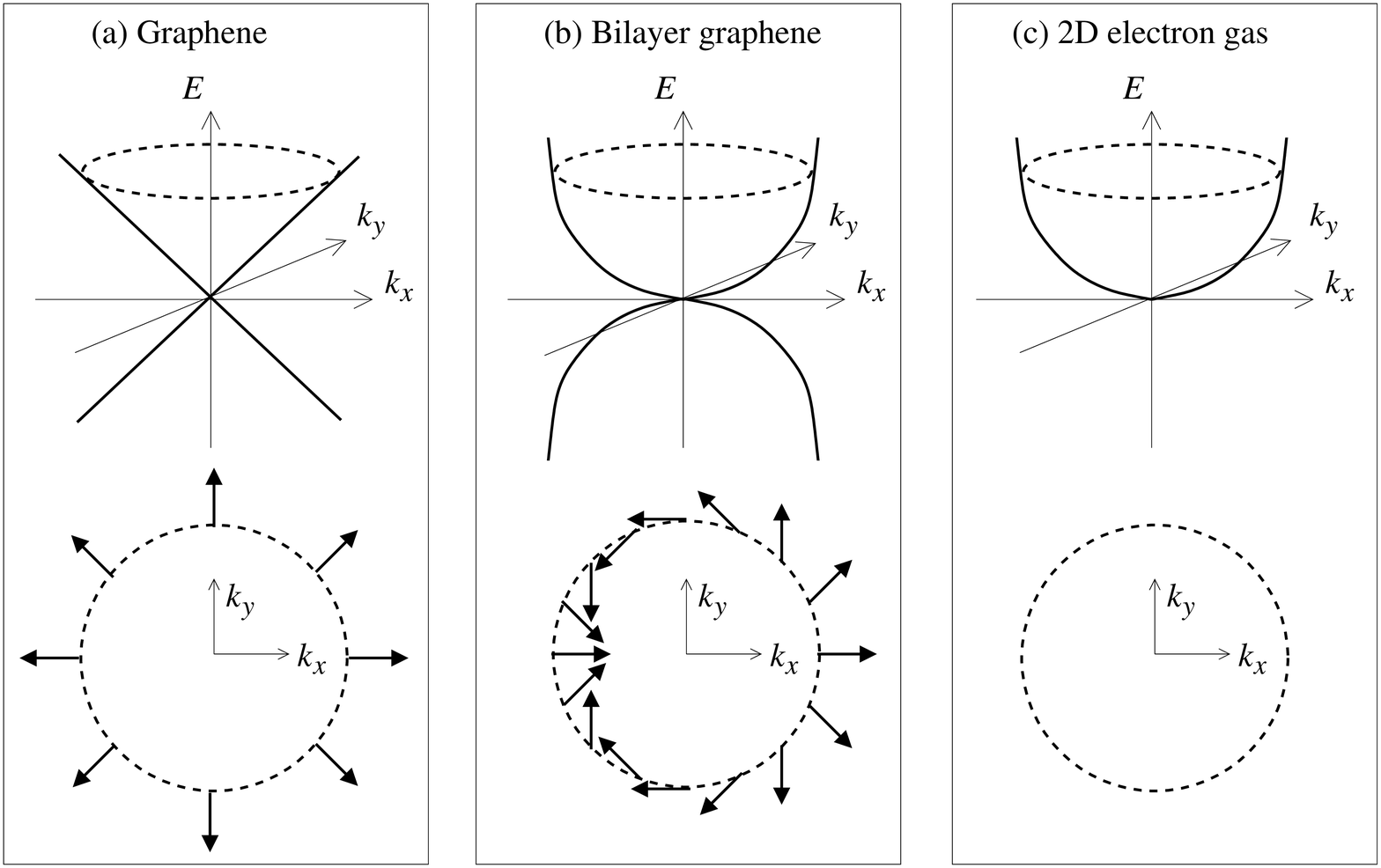}
  \caption{(a) Upper panel: Energy dispersions
  for low-energy electrons
  in graphene. Lower panel: Pseudospin distribution for electronic
  eigenstates in graphene
  on an equi-energy contour specified by the dashed curve in the upper
  panel.  The arrows represent the direction of pseudospin.
  Here, we consider the electronic states with wavevectors
  near the Dirac point K.
  (b) and (c): Similar quantities as in (a) for
  bilayer graphene and a conventional 2DEG, respectively.
  Note that the electrons in a 2DEG
  have an energy minimum at the center of the Brillouin zone and
  do not have a pseudospin degree of freedom.
  }
  \label{Fig2}
  \end{figure*}

When a second layer is added --- thus forming bilayer
graphene [Fig.~1(a)] --- the electronic structure changes dramatically.
The dispersion for low-energy quasiparticles
in graphene is linear [Fig.~2(a)],
whereas in bilayer graphene it becomes quadratic [Fig.~2(b)].
The quantum Hall conductance in bilayer graphene has been theoretically
predicted~\cite{PhysRevLett.96.086805}, and experimentally
verified (again by using mechanically exfoliated
samples)~\cite{novoselov_unconventional_2006}
to be $\sigma_{xy}=n\, e^2/h$ per spin and valley degrees
of freedom. This is very similar to the case of a conventional 2DEG, but, very significantly, with
no plateau at $n=0$.
This novel quantum Hall effect in bilayer graphene
has been explained by the appearance of a non-trivial Berry phase of $2\pi$ in
the electronic wavefunction~\cite{PhysRevLett.96.086805,novoselov_unconventional_2006},
different from the case of a conventional 2DEG.
These pioneering studies~\cite{PhysRevLett.96.086805,novoselov_unconventional_2006}
have significantly influenced
the community working on the novel physics of graphene nanostructures, and
the concept of a non-trivial Berry phase in the electronic wavefunction
of bilayer graphene has been commonly accepted thereafter.

  \begin{figure}
  \includegraphics[width=1.0\columnwidth]{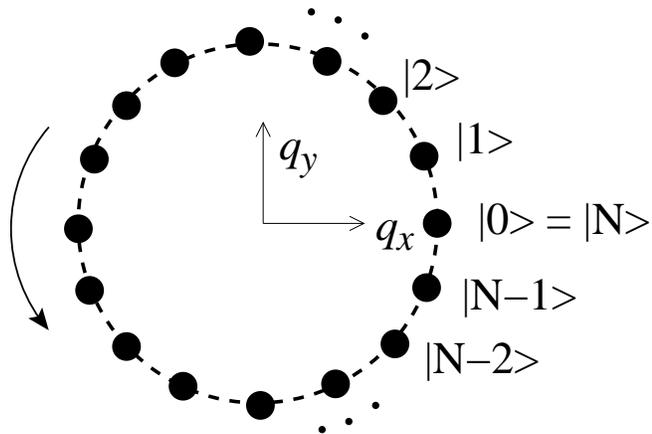}
  \caption{Schematics of electronic eigenstates
  in graphene or multi-layer graphene on an equi-energy
  contour with a wavevector around the Dirac point K.
  We define ${\bf q}={\bf k}-{\bf K}$ where {\bf k} is the
  2D Bloch wavevector.
  If we start from the state $\left|0\right>$, we end up with
  the state $\left|N\right>$ after a complete
  counterclockwise rotation along the contour
  ($\left|N\right>=\left|0\right>$).  Here,
  the evolving parameter is ${\bf q}$.}
  \label{Fig3}
  \end{figure}

In this paper, we show that the Berry phase in the electronic
wavefunction of bilayer graphene is equivalent to that of a conventional
2DEG
{(i.\,e.\,, Berry phase of $2\pi$, $4\pi$, etc., are
{\it not} non-trivial but are all equivalent to 0)}
by adopting simple arguments on the gauge
freedom of electronic wavefunctions.
We then reinterpret the original experimental
results~\cite{novoselov_unconventional_2006}:
{although the quantum Hall effect in bilayer graphene is qualitatively
different from that in a conventional 2DEG, what we need for an explanation
of that novel phenomenon is
not a non-trivial Berry phase of $2\pi$ but the concept of}
{\it pseudospin winding number}, i.\,e.\,,
how many times the pseudospin vector
-- the vector representing the
relative phase on the two different sublattices of carbon
atoms~\cite{PhysRev.71.622}
[see lower panels of Figs.~2(a) and~2(b),
and Eqs.~(\ref{eq:wfn_mono}) and~(\ref{eq:wfn_bi})] --
rotates when the electronic wavevector undergoes one
full rotation around the Dirac point [Figs.~1(b) and~3].
We note that this pseudospin winding number is exactly the
same quantity as the `degree of chirality' $J$ of previous
studies~\cite{novoselov_unconventional_2006,PhysRevLett.96.086805}.
On the other hand, the term `chirality' -- in connection with
neutrino physics -- has been used to refer to
the inner product between the two-dimensional (2D)
Bloch wavevector from the Dirac point and the pseudospin
vector~\cite{RevModPhys.81.109}; hence,
for graphene, the chiralities for electronic states with wavevector
close to the K point in the upper band and in the lower band
would be $+1$ and $-1$, respectively, whereas the
degree of chirality is $+1$ regardless of the band index.
To avoid this confusion, we suggest the term `pseudospin
winding number' here.  Also, pseudospin winding number has been
defined in the form of a more general topological
invariant~\cite{heikkila_volovik,volovik,PhysRevB.66.235110}.

\section{II. Berry phase}

For electrons in a periodic system, the Berry phase is a phase acquired by the wavefunction
over the course of a cyclic evolution of the Hamiltonian - such concept has had
a fundamental role in developing the modern theory of polarization~\cite{PhysRevB.47.1651,RevModPhys.66.899}
and of magnetization~\cite{PhysRevLett.95.137205,PhysRevB.74.024408}.

The evolving parameter that we consider here is the 2D Bloch wavevector
${\bf k}=(k_x,k_y)$.
The Berry phase $\Gamma$ described in the wavevector
evolution of Fig.~3 is given by
\begin{equation}
\Gamma=-i\lim_{N\to\infty}\sum_{j=0}^{N-1}\log\left<j|j+1\right>\,,
\label{eq:berry2}
\end{equation}
where
\begin{equation}
\left|N\right>=\left|0\right>\,.
\label{eq:0_N}
\end{equation}
Here, {\bf k} is determined by the state index {\it j}.

{We first show that Berry phase [defined in
Eq.~(\ref{eq:berry2})] is in general meaningful only modulo $2\pi$.
After multiplying each of the wavefunctions by a
{position-independent} overall phase factor $e^{i\,\theta_j}$,
we obtain another set of wavefunctions which satisfies the
equation of motion
\begin{equation}
\left|j\right>'=e^{i\,\theta_j}\,\left|j\right>\,.
\label{eq:j2jp}
\end{equation}
In order to have a well-defined set of wavefunctions
for Berry phase evaluation
[Eq.~(\ref{eq:0_N})], we have to impose
\begin{equation}
\theta_N-\theta_0=2\,\pi\,m
\label{eq:N_0_m}
\end{equation}
where $m$ is an arbitrary integer.
Now using Eqs.~(\ref{eq:berry2}), (\ref{eq:j2jp}),
and~(\ref{eq:N_0_m}), the Berry phase $\Gamma'$ for the
new set of wavefunctions [Eq.~(\ref{eq:j2jp})] is
\begin{eqnarray}
\Gamma'&=&-i\lim_{N\to\infty}\sum_{j=0}^{N-1}\log\left<j|j+1\right>'\nonumber\\
&=&\Gamma+\lim_{N\to\infty}\sum_{j=0}^{N-1}\left(\theta_{j+1}-\theta_j\right)\nonumber\\
&=&\Gamma+2\pi\,m\,,
\label{eq:berry2p}
\end{eqnarray}
i.\,e.\,, Berry phase can be defined modulo $2\pi$ only.
In the following, we paraphrase this general argument
using the language of graphene.
}

For graphene, let us consider the case where {\bf k}
is very close to the Dirac point K [Fig.~1(b)],
and define ${\bf q}={\bf k}-{\bf K}$ ($|{\bf q}|\ll|{\bf K}|$).
Then, if we use a basis set composed of the Bloch sums of $p_z$
orbitals localized on the two sublattices A and B [Fig.~1(a)],
the effective Hamiltonian reads~\cite{PhysRev.71.622}
\begin{equation}
H_{\rm mono}({\bf q})=\hbar\, v_0\,q\left(
\begin{array}{cc}
0 & \exp(-i\theta_{\bf q})\\
\\
\exp(i\theta_{\bf q}) & 0\\
\end{array}
\right)\,,
\label{eq:H_mono}
\end{equation}
where $v_0$ is the band velocity and $\theta_{\bf q}$
the angle between {\bf q} and the $+k_x$ direction.
The energy eigenvalue and
wavefunction of Eq.~(\ref{eq:H_mono})
are given by $E^{\rm mono}_{s\,{\bf q}}=\hbar\,v_0\,s\,|{\bf q}|$
and
\begin{equation}
\left|\psi^{\rm mono}_{s\,{\bf q}}\right>=\frac{1}{\sqrt{2}}
\left(
\begin{array}{c}
1
\\
s\,e^{i\theta_{\bf q}}
\end{array}
\right)\,,
\label{eq:wfn_mono}
\end{equation}
respectively, where $s=\pm1$ is the band index.
As shown before~\cite{ando_berrys_1998},
the Berry phase $\Gamma_{\rm mono}$ of the electron
wavefunction in graphene when
the Bloch wavevector undergoes a full rotation around the
Dirac point (Fig.~3) can be obtained from Eq.~(\ref{eq:berry2}):
\begin{eqnarray}
\Gamma_{\rm mono}&=&-i\lim_{N\to\infty}\sum_{j=0}^{N-1}\log
\left(\frac{1+\exp\left[i(\theta_{j+1}-\theta_j)\right]}{2}\right)
\nonumber\\
&=&
-i\int_0^{2\pi}d\theta\,\frac{i}{2}\nonumber\\
&=&\pi\,,
\label{eq:berry_mono}
\end{eqnarray}
where we have substituted {\bf q} by the corresponding
state index {\it j} ($j=0,\,1,\,2,\,\cdots,\,N-1$) as shown in Fig.~3.

Now we show that Eq.~(\ref{eq:berry2}) is not affected
by an arbitrary phase factor of each electronic wavefunction.
Suppose that we use a new form for an electron wavefunction
in graphene
\begin{eqnarray}
\left|\psi^{\rm mono}_{s,j}\right>'
&=&e^{-i\theta_j/2}\,\left|\psi^{\rm mono}_{s,j}\right>
\nonumber\\
&=&\frac{1}{\sqrt{2}}
\left(
\begin{array}{c}
e^{-i\theta_j/2}
\\
s\,e^{i\theta_j/2}
\end{array}
\right)\,,
\label{wfn_mono2}
\end{eqnarray}
for $j=0,1,...,N-1$, and of course
$\left|\psi^{\rm mono}_{s,N}\right>'=\left|\psi^{\rm mono}_{s,0}\right>'$
(Fig.~3).
Then the Berry phase $\Gamma'_{\rm mono}$ is given by
\begin{eqnarray}
\Gamma'_{\rm mono}&=&-i\lim_{N\to\infty}
\{\log\left<\psi^{\rm mono}_{s,N-1}|
\psi^{\rm mono}_{s,0}\right>'\nonumber\\
&&+\sum_{j=0}^{N-2}\log
\left(\frac{
e^{-i(\theta_{j+1}-\theta_j)/2}+
e^{i(\theta_{j+1}-\theta_j)/2}}{2}\right)\}
\nonumber\\
&=&-i\left\{
i\,\pi+
\int_0^{2\pi}d\theta\,0
\right\}
\nonumber\\
&=&\pi\,,
\label{eq:berry_mono2}
\end{eqnarray}
i.\,e.\,, we obtain the same result
as in Eq.~(\ref{eq:berry_mono}).
The physical reason for this invariance of the
Berry phase with respect to an arbitrary gauge
phase of each electronic state
evaluated from Eq.~(\ref{eq:berry2})
is that each state appears twice
(once as a bra state and once as a ket state), thus
cancelling out any arbitrary phase in each wavefunction.
Since the Berry phase obtained from
Eq.~(\ref{eq:berry2}) is {\it not} affected by
any arbitrary phase of each state $\left|j\right>$,
we can choose for convenience a gauge such that the state
varies smoothly over the course of parameter evolution.

Similarly, for bilayer graphene,
if we use a basis set composed of Bloch sums of localized Wannier-like
$p_z$ orbitals on each of the two sublattices A and B$'$
[Fig.~1(a)]~\cite{PhysRevLett.95.076804},
and consider only the nearest-neighbor intra-layer hopping
and vertical interlayer hopping
(characterized by the hopping integral $t_\perp$) processes,
the effective Hamiltonian of low-energy electronic states
in bilayer graphene becomes~\cite{novoselov_unconventional_2006}
\begin{equation}
H_{\rm bi}({\bf q})=-\frac{\hbar^2\,q^2}{2m^*}\left(
\begin{array}{cc}
0 & \exp(-2i\theta_{\bf q})\\
\\
\exp(2i\theta_{\bf q}) & 0\\
\end{array}
\right)\,,
\label{eq:H_bi}
\end{equation}
where $m^*=3\,a^2\,t_\perp/8v_0^2$ is the effective mass
and $a$ the lattice parameter.
The energy eigenvalue and
wavefunction of Eq.~(\ref{eq:H_bi})
are given by $E^{\rm bi}_{s\,{\bf q}}=s\,{\hbar^2\,q^2}/{2m^*}$
and
\begin{equation}
\left|\psi^{\rm bi}_{s\,{\bf q}}\right>=\frac{1}{\sqrt{2}}
\left(
\begin{array}{c}
1
\\
-s\,e^{2i\theta_{\bf q}}
\end{array}
\right)\,,
\label{eq:wfn_bi}
\end{equation}
respectively ($s=\pm1$).
The Berry phase $\Gamma_{\rm bi}$ of the electron
wavefunction in bilayer graphene
can again be obtained from Eq.~(\ref{eq:berry2}):
\begin{eqnarray}
\Gamma_{\rm bi}&=&-i\lim_{N\to\infty}\sum_{j=0}^{N-1}\log
\left(\frac{1+\exp\left[2i(\theta_{j+1}-\theta_j)\right]}{2}\right)
\nonumber\\
&=&
-i\int_0^{2\pi}d\theta\,i\nonumber\\
&=&2\pi\,,
\label{eq:berry_bi}
\end{eqnarray}
which is the result of previous
studies~\cite{PhysRevLett.96.086805,novoselov_unconventional_2006}.

Now we consider another form for the wavefunction
\begin{equation}
\left|\psi^{\rm bi}_{s\,{\bf q}}\right>'=\frac{1}{\sqrt{2}}
\left(
\begin{array}{c}
e^{-i\theta_{\bf q}}
\\
-s\,e^{i\theta_{\bf q}}
\end{array}
\right)\,,
\label{eq:wfn_bi2}
\end{equation}
whose only difference from the original one
is the {position-independent} overall phase
($\left|\psi^{\rm bi}_{s\,{\bf q}}\right>'
=e^{-i\theta_{\bf q}}\,\left|\psi^{\rm bi}_{s\,{\bf q}}\right>$).
{Because the two wavefunctions are different only
in the overall coefficient, they both are perfectly good
solutions of the effective Hamiltonian of bilayer graphene
[Eq.~(\ref{eq:H_bi})]].}
Moreover, both the original [Eq.~(\ref{eq:wfn_bi})]
and new [Eq.~(\ref{eq:wfn_bi2})] wavefunctions
satisfy the gauge condition that the wavefunction varies
smoothly in the course of cyclic evolution;
{hence, in the evaluation of the Berry phase,
one does not have to consider any branch-cut as
we did in the derivation of Eq.~(\ref{eq:berry_mono2})}.
The Berry phase $\Gamma'_{\rm bi}$ for
 $\left|\psi^{\rm bi}_{s\,{\bf q}}\right>'$ evaluated
from Eq.~(\ref{eq:berry2}) is
\begin{equation}
\Gamma'_{\rm bi}=0\,;
\label{eq:berry_bi2}
\end{equation}
in other words, {\it the Berry phase of the electronic wavefunction
of bilayer graphene is the same as that of a conventional 2DEG.}

In fact, the Berry phase of an electronic wavefunction
in bilayer graphene evaluated from Eq.~(\ref{eq:berry2})
is $2\,m\,\pi$, with $m$ an arbitrary integer.
{Or, in other words, any Berry phase of $2\,m\,\pi$ is equivalent
to a trivial Berry phase of 0.}
[Similarly, this Berry phase for graphene obtained from Eq.~(\ref{eq:berry2})
is $(2m+1)\,\pi$, with $m$ an arbitrary integer.]
This statement is related also to the well known application of
Berry-phase physics to the modern theory of
polarization~\cite{PhysRevB.47.1651,RevModPhys.66.899}, according to which
the electrical polarization in a periodic system can only
be determined modulo $e\,{\bf R}/\Omega$, where {\bf R} is a
lattice vector and $\Omega$ the unit cell volume.

{Our claim that the Berry phase in bilayer graphene is {\it not} non-trivial and
is equivalent to that of a conventional 2DEG remains valid even if we consider
a more complex model Hamiltonian.
If, for example, second-nearest-neighbor
inter-layer hopping processes are considered in the
tight-binding model of bilayer graphene, the low-energy
electronic states can be described by four Dirac
cones~\cite{PhysRevLett.96.086805}.
It was suggested that the central Dirac cone contributes
$-\pi$ to the Berry phase and each of the three satellite Dirac cones contribute
$\pi$ to the Berry phase and hence the total ``non-trivial'' Berry phase is
$-\pi+3\times\pi=2\pi$~\cite{PhysRevLett.96.086805}.
According to our discussion, this view is not correct, again because
a Berry phase of $-\pi$ is equivalent to that of $\pi$, etc.  In particular,
starting from the wavefunctions used in Ref.~\cite{PhysRevLett.96.086805},
one can define a new set of wavefunctions with additional four phase factors,
e.\,g.\,, $e^{-i\,\theta_{\bf q}}$, defined around each of the four
Dirac cones in a way similar to Eq.~(\ref{eq:wfn_bi2}).
Then, the Berry phase around the $i^{\rm th}$
Dirac cone ($i=1,2,3,4$) obtained by evaluating Eq.~(\ref{eq:berry2}) would be
$(2m_i+1)\pi$ for an arbitrary integer $m_i$.
Thus, the Berry phase for bilayer graphene is of the form
$\Gamma_{\rm bi}=\sum_{i=1}^4(2m_i+1)\pi=2m\pi$,
with $m$ an arbitrary integer, and cannot uniquely be
$2\pi$ as suggested in previous
studies~\cite{novoselov_unconventional_2006,PhysRevLett.96.086805}.}

\section{III. Pseudospin winding number}

However, these considerations do not mean that there is no qualitative difference
between the electronic eigenstates in bilayer graphene and
those in a conventional 2DEG.
Indeed, a new phenomenon in bilayer graphene was
observed~\cite{novoselov_unconventional_2006}
and explained~\cite{PhysRevLett.96.086805}:
the step size at charge neutrality point in the quantum Hall
conductance $\sigma_{xy}$ in bilayer graphene
is $2e^2/h$ per spin and valley degrees of freedom
and not $e^2/h$ as in a conventional 2DEG.
We reinterpret this result as a direct
manifestation of the {\it pseudospin winding number},
and explain this in the following.

For an electronic state whose wavefunction can be defined as
a two-component spinor, as in graphene [Eq.~(\ref{eq:wfn_mono})]
or in bilayer graphene [Eq.~(\ref{eq:wfn_bi})], a 2D pseudospin
vector can be defined from the relative phase of the two
components~\cite{PhysRev.71.622,novoselov_unconventional_2006}.
Figures~2(a) and~2(b) show the pseudospin vector for electronic
wavefunctions in graphene and in bilayer graphene, respectively.
The pseudospin winding number $n_w$ is then defined as the number of
rotations that a pseudospin vector undergoes
when the electronic wavevector rotates fully one time
around the Dirac point:
it is immediate to see from Figs.~2(a) and~2(b) that
$n_w$ is 1 in graphene and 2 in bilayer graphene.

The low-energy effective Hamiltonian of a graphene multilayer,
with {\it n} layers and ABC stacking, is~\cite{PhysRevB.77.155416}
\begin{equation}
H_n({\bf q})\propto q^n\left(
\begin{array}{cc}
0 & \exp(-n\,i\,\theta_{\bf q})\\
\\
\exp(n\,i\,\theta_{\bf q}) & 0\\
\end{array}
\right)\,.
\label{eq:Heff_n}
\end{equation}
Eigenvalues and eigenfunctions of Eq.~(\ref{eq:Heff_n})
are given by $E^{n}_{s\,{\bf q}}\propto s\,q^n$
and
\begin{equation}
\left|\psi^{n}_{s\,{\bf q}}\right>=\frac{1}{\sqrt{2}}
\left(
\begin{array}{c}
1
\\
-s\,e^{n\,i\,\theta_{\bf q}}
\end{array}
\right)
\label{eq:wfn_n}
\end{equation}
respectively ($s=\pm1$).
Thus, the pseudospin winding number $n_w$ of the eigenstate
in Eq.~(\ref{eq:Heff_n}) is $n$.

It has been shown
(Supplementary Information of Ref.~\cite{novoselov_unconventional_2006})
that a system described by Eq.~(\ref{eq:Heff_n})
has a quantum Hall conductance step at the charge neutrality point
of size $n\,e^2/h$ per spin and valley degrees of freedom.
Our interpretation is that what we learn from a measured step
size $n\,e^2/h$ per spin and valley degrees of
freedom~\cite{novoselov_unconventional_2006}
is not the ``absolute'' value of the Berry phase
($\Gamma=n\pi$)
but the pseudospin winding number ($n_w=n$);
We also note that the quantum Hall conductance step, which has been correctly
obtained in Ref.~\cite{novoselov_unconventional_2006} from
Eq.~(\ref{eq:Heff_n}), does not need a Berry phase of $\Gamma=n\pi$
for its explanation.
In other words, the size of the quantum Hall conductance step at the charge
neutrality point is connected directly to {the special form of the
Hamiltonian in Eq.~(\ref{eq:Heff_n}), or, equivalently, to $n$ which
can be interpreted, from its wavefunction [Eq.~(\ref{eq:wfn_n})],
as} the pseudospin winding number
and not to the absolute value of the Berry phase.
{(It should be stressed that our discussion of the pseudospin winding number
is confined to systems whose effective Hamiltonian is given by
Eq.~(\ref{eq:Heff_n}).)}

On the other hand, the Berry phase still determines the shift of the plateaus,
with integer and half-integer quantum Hall effects for $\Gamma=0$ and $\Gamma=\pi$,
respectively~\cite{ando_berrys_1998,novoselov:2005Nat_Graphene_QHE,zhang:2005Nat_Graphene_QHE}.
The former and the latter correspond
to multi-layer graphene with even and odd number of layers,
respectively~\cite{PhysRevB.77.155416}.  It has been recently shown
within a semiclassical theory that the so-called topological part of
the Berry phase, which is determined by the pseudospin winding number,
is responsible for this shift~\cite{fuchs_topological_2010}.

\section{IV. Conclusion}

In conclusion, we have shown that the Berry phase of the low-energy quasiparticle
wavefunction of bilayer graphene is the same
as that of a conventional two-dimensional electron gas,
and that the fundamental difference between the two systems is
the pseudospin winding number.
{Our findings have a broader implication than
the cases discussed here.
For example, the interpretation of
the recent angle-resolved photoemission spectroscopy
on bilayer graphene~\cite{PhysRevB.84.125422,PhysRevLett.107.166803}
and that of the recent quantum Hall
experiments on trilayer
graphene~\cite{trilayer_natphys} should be revisited in view
of the present discussion.}

\section{Acknowledgments}

We thank Steven G. Louie, 
Alessandra Lanzara, Choonkyu Hwang, and Davide Ceresoli
for fruitful discussions, and
Intel Corporation for support (CHP).

\end{document}